\documentclass[12pt,preprint]{aastex}

\usepackage{ragged2e}
\justifying\let\raggedright\justifying

\usepackage{threeparttable}

\usepackage{graphicx}
\usepackage{epsfig}
\usepackage{bm}
\usepackage{subfigure}
\usepackage{multicol}
\usepackage{multirow}
\usepackage{array}
\newcommand{\PreserveBackslash}[1]{\let\temp=\\#1\let\\=\temp}
\newcolumntype{C}[1]{>{\PreserveBackslash\centering}p{#1}}
\newcolumntype{R}[1]{>{\PreserveBackslash\raggedleft}p{#1}}
\newcolumntype{L}[1]{>{\PreserveBackslash\raggedright}p{#1}}

\shortauthors{Song et. al} \shorttitle{White-light Flare by Reconnection}

\begin{document}

\title{A White-light Flare Powered by Magnetic Reconnection in the Lower Solar Atmosphere }

\author{Yongliang Song\altaffilmark{1,2}, Hui Tian\altaffilmark{2,1}, Xiaoshuai Zhu\altaffilmark{3}, Yajie Chen\altaffilmark{2,3}, Mei Zhang\altaffilmark{1,4}, Jingwen Zhang\altaffilmark{5}}

\altaffiltext{1}{Key Laboratory of Solar Activity, National Astronomical Observatories, Chinese Academy of Sciences, Beijing 100012, People's Republic of China; \\ Email: ylsong@bao.ac.cn}
\altaffiltext{2}{School of Earth and Space Sciences, Peking University, Beijing 100871, People's Republic of China; \\ Email: huitian@pku.edu.cn}
\altaffiltext{3}{Max Planck Institute for Solar System Research, Justus-von-Liebig-Weg 3, D-37077 GoÃÂttingen, Germany}
\altaffiltext{4}{School of Astronomy and Space Science, University of Chinese Academy of Sciences, Beijing 100049, People's Republic of China}
\altaffiltext{5}{Institute for Astronomy, University of Hawaii, Honolulu, HI 96822-1897, USA.}

\begin{abstract}
White-light flares (WLFs), first observed in 1859, refer to a type of solar flares showing an obvious enhancement of the visible continuum emission. This type of enhancement often occurs in most energetic flares, and is usually interpreted as a consequence of efficient heating in the lower solar atmosphere through non-thermal electrons propagating downward from the energy release site in the corona. However, this coronal-reconnection model has difficulty in explaining the recently discovered small WLFs. Here we report a C2.3 white-light flare, which are associated with several observational phenomena: fast decrease in opposite-polarity photospheric magnetic fluxes, disappearance of two adjacent pores, significant heating of the lower chromosphere, negligible increase of hard X-ray flux, and an associated U-shaped magnetic field configuration. All these suggest that this white-light flare is powered by magnetic reconnection in the lower part of the solar atmosphere rather than by reconnection higher up in the corona.
\end{abstract}



\section{Introduction}

White-light flares (WLFs), first observed in 1859 by Richard C. Carrington (Carrington 1859) and Richard Hodgson (Hodgson 1859), refer to a type of solar flares showing an obvious enhancement of the visible continuum emission. Though more than one and half centuries have passed, the generation mechanism of this interesting and important phenomenon still remains a challenging topic in solar physics (Hudson 2016).

The white-light emission is likely related to hydrogen atom recombination or/and negative hydrogen emission, and its altitude is generally believed to be the lower chromosphere or/and upper photosphere (e.g., Aboudarham \& Henoux 1989; Ding et al. 1994; Fang \& Ding 1995). However, it is still unclear how the lower atmosphere is heated to produce the white-light emission. For a long time, it was believed that only the most energetic flares are able to deposit sufficient energy to the deep solar atmosphere to produce the white-light continuum emission. This big flare syndrome (Kahler 1982) was questioned after some weak C-class WLFs were detected (e.g., Matthews et al. 2003; Hudson et al. 2006; Jess et al. 2008). The finding of small WLFs appears to provide support to the suggestion that all flares are WLFs, which attributes the nondetectability of many small WLFs to the limitation of the telescope sensitivity and resolution (Neidig 1989; Hudson 2006; Jess et al. 2008). However, this idea is highly controversial, and was not supported by recent high-resolution observations (Yurchyshyn et al. 2017) from the Goode Solar Telescope (GST; Cao et al. 2010). 

Some WLFs are found to show a high temporal and spatial correlation between the white-light enhancement and the hard X-ray emission (e.g., Fletcher et al. 2007; Watanabe et al. 2010; Hao et al. 2012; Kuhar et al. 2016; Song et al. 2018a), suggesting that non-thermal electrons might play an important role in the generation of the white-light emission. Other WLFs are not cospatial with hard X-ray sources, or not accompanied by obvious hard X-ray emission at all (e.g., Ryan et al. 1983; Ding et al. 1994; Sylwester \& Sylwester 2000). It is still unclear how the enhancement of white-light emission is produced in these WLFs. There is one suggestion that these WLFs might be powered by direct energy release in the lower atmosphere (e.g., Ding et al. 1994; Chen et al. 2001). However, no direct observational evidence has been found to support this hypothesis. 

Using high-quality observations from several solar telescopes, we have conducted a detailed investigation of a small C2.3 WLF. Our results strongly suggest that magnetic reconnection in the lower atmosphere directly powers the white-light emission in this small flare.


\section{Observations}

This flare occurred on 2016 November 30 in active region (AR) NOAA 12615, which was still in its emerging phase from the solar interior to the atmosphere. Several space-borne observatories, including the Solar Dynamics Observatory (SDO; Pesnell et al. 2012), the Interface Region Imaging Spectrograph (IRIS; De Pontieu et al. 2014), the FERMI Gamma-Ray Space Telescope and the Geostationary Orbiting Environmental Satellites (GOES), have recorded this flare. The flare was observed as a compact transient brightening, lasting approximately from 15:22 UT to 15:38 UT, in the ultraviolet (UV) and extreme-ultraviolet (EUV) images taken by IRIS and the Atmospheric Imaging Assembly (AIA; Lemen et al. 2012) on board SDO (Fig.1a). The 1--8 \AA~soft X-ray flux measured by GOES peaked around 15:24 UT, and the peak flux indicated a flare class of C2.3 (Fig.1c).

The AIA instrument observes the full solar disk in seven EUV passbands (94 \AA, 131 \AA, 171 \AA, 193 \AA, 211 \AA, 304 \AA, 335 \AA) and two UV passbands (1600 \AA, 1700 \AA). These passbands sample plasma with different temperatures in the solar atmosphere, with a spatial pixel size of $\sim0.6^{\prime\prime}$. The time cadences for the EUV and UV observations are 12 s and 24 s, respectively. Here we only used images in the 94 \AA~and 1700 \AA~passbands. The Helioseismic and Magnetic Imager (HMI; Scherrer et al. 2012) on board SDO performs spectropolarimetric measurements of the photosphere at six wavelength positions across the spectral profile of the Fe {\footnotesize I}  6173 \AA line. Photospheric continuum intensity images are reconstructed from the six-point Stokes-I profiles (Couvidat et al. 2012). In addition, photospheric magnetograms are obtained through inversion of the full set of Stokes parameters. The time cadences for the HMI continuum images, Dopplergrams, line-of-sight (LOS) magnetograms and vector magnetograms are 45 sec, 45 sec, 45 sec and 12 min, respectively. The spatial pixel size for the HMI data is about $\sim0.5^{\prime\prime}$. We used the AIA and HMI data of NOAA AR 12615 from 15:00 UT to 16:00 UT on 2016 November 30. The AIA and HMI images were calibrated using the standard aia\_prep.pro routines in SolarSoft (SSW).

IRIS observed NOAA AR 12615 from 14:39 UT to 15:40 UT with a medium coarse 8-step raster ($60^{\prime\prime}$ along the slit, 8 raster steps with a $2^{\prime\prime}$ step size). Slit-jaw images (SJIs) in the filters of 1330 \AA~and 2832 \AA, which sample mainly emission of the strong C {\footnotesize II} 1334.53/1335.71\AA~spectral lines formed around 30,000 Kelvin and the Mg {\footnotesize II} wing emission around 2832 \AA~formed in the upper photosphere respectively, were also taken at a time cadence of ~18 sec and a spatial pixel size of $\sim0.33^{\prime\prime}$ during this period (Fig 1a). We used the calibrated level-2 data of IRIS, in which the dark-current subtraction, flat fielding, and geometrical correction have all been taken into account (De Pontieu et al. 2014). The AIA and IRIS data were co-aligned by matching commonly observed features in both the AIA 1700 \AA~ and SJI 1330 \AA~images. 

The Fermi Gamma-Ray Space Telescope is a NASA mission designed to explore the high-energy phenomena in the universe. There are two instruments on board this space telescope, the Large Area Telescope (LAT; Atwood et al. 2009) and the Gamma-ray Burst Monitor (GBM; Meegan et al. 2009). The GBM is sensitive to X-rays and gamma rays with energies between 8 keV and 40 MeV. Fermi/GBM can detect solar flares with six sun-facing detectors (NaI 0--5). Flares recorded by Fermi/GBM can be found on the following website: https://hesperia.gsfc.nasa.gov/fermi/gbm/qlook/fermi\_gbm\_flare\_list.txt. The energy spectrum of Fermi for the flare studied here was generated by the NaI 5 detector, which was almost exactly sun-directed during the occurrence of the flare. 

\section{Results and discussion}

Observations from SDO/HMI revealed an enhancement of the visible continuum (white-light) emission in this flare (Fig.1a). The maximum enhancement of continuum intensity with respect to the pre-flare intensity is about 18\% (Fig.1b). It should be noted that an HMI continuum intensity image is reconstructed from a set of filtergrams taken at six wavelength positions across the Fe {\footnotesize I} 6173.3 \AA~absorption line, and that sometimes line core emission could lead to a false signal of enhanced continuum ({\v{S}}vanda et al. 2018). We thus examined the six-point line profiles observed at four different times at nine spatial pixels in the region of white-light enhancement, and found an obvious continuum enhancement around the flare peak time (Fig.1b, d). This result demonstrates that this flare is a typical WLF. 

For some WLFs, the white-light enhancement is known to show a high temporal and spatial correlation with the hard X-ray emission (e.g., Hao et al. 2012; Kuhar et al. 2016), indicating a close relationship between the white-light enhancement and accelerated electrons. These WLFs are often accompanied by strong emission of hydrogen Balmer lines and a strong Balmer jump in the spectra. They are classified as type I WLFs (Fang \& Ding 1995) and are thought to be related to high-energy non-thermal electrons propagating downward from the site of magnetic reconnection in the corona (Aboudarham et al. 1986; Machado et al. 1989; Metcalf et al. 2003). These non-thermal electrons may be stopped in the upper chromosphere, and then produce enhanced continuum emission by back-warming the lower atmosphere (photosphere or lower chromosphere) through UV radiation. Alternatively, these electrons might penetrate the lower chromosphere or even the photosphere and directly heat the local medium, producing enhanced continuum emission. However, theoretical investigations suggest that a minimum energy of 100 keV (Aboudarham et al. 1986) is required for the electrons to penetrate the lower chromosphere, and 900 keV to penetrate the photosphere (Neidig 1989). From observation of the FERMI/GBM, we see an obvious enhancement of the 4-15 keV and 15-25 keV X-ray flux at the time of enhanced white-light emission. However, there is only a marginal enhancement of the 25-50 keV hard X-ray emission and no hard X-ray emission above 50 keV during this WLF (Fig. 2a), suggesting that direct heating by non-thermal electrons transported from a coronal reconnection site may not be able to explain the generation of this WLF.

Some other WLFs are characterized by a weak or no correlation with hard X-ray emission (e.g., Ryan et al. 1983; Sylwester \& Sylwester 2000), weak Balmer line emission and no obvious Balmer jump. They are classified as type II WLFs (Fang \& Ding 1995), and their heating mechanisms are poorly understood. There is one suggestion that the energy release site might be in the lower atmosphere rather than in the corona for these WLFs (e.g., Ding et al. 1994; Chen et al. 2001). Without Balmer line observations, it is unclear whether our observed WLF is a type II WLF. However, as described in the following, our observations do provide strong evidence to support the scenario of energy release through magnetic reconnection in the lower atmosphere.

Photospheric magnetic field observations from SDO/HMI revealed a connection between this WLF and interaction of opposite-polarity magnetic fields. The region of white-light enhancement is just located at the interface between two small sunspots (pores) with different magnetic polarities (Fig. 2b). Interaction between the two led to a fast decrease in the magnetic flux of either polarity, suggesting the occurrence of magnetic reconnection (e.g., Wang \& Shi 1993; Yan et al. 2016) (Fig. 2c, e). The HMI Doppolergrams shown in Fig. 2d reveal predominant blue shift, clearly related to the emergence of the active region. The magnitude of blue shift around the polarity inversion line (PIL) obviously decreased when flux cancellation started, which is likely caused by the partial compensation of the emerging motion by the reconnection downflow. The WLF occurred synchronously to this flux cancellation process. The spatial-temporal correlation between flux cancellation and this WLF strongly suggests that this WLF is powered by magnetic reconnection in the lower solar atmosphere (e.g., Syntelis et al. 2019). The two small sunspots disappeared about half an hour after the white-light enhancement, likely resulted from a reconfiguration of the magnetic field structure due to the reconnection.  

With IRIS observations, we found that this WLF can also be identified as a UV burst (Peter et al. 2014; Young et al. 2018). UV bursts are characterized by the superposition of chromospheric absorption lines (e.g., Ni {\footnotesize II} 1393.33 \AA~and 1335.20 \AA) on greatly enhanced and broadened spectral profiles of several emission lines formed at temperatures of a few tens of thousand Kelvin (e.g., Si {\footnotesize IV} 1393.76/1402.77 \AA, C {\footnotesize II} 1334.53/1335.71 \AA) (Fig. 3), indicating local plasma heating in the lower atmosphere during the process of large-scale magnetic flux emergence (e.g., Tian et al. 2018). The accompanied flux cancellation suggests that the heating is most likely caused by magnetic reconnection. Schmieder et al. (2004) found local plasma heating to coronal temperatures by low-altitude reconnection in an extraordinarily intense Ellerman bomb. Without H$\alpha$ wing observations it is unclear whether our event falls into the category of Ellerman bombs, though it is known that a large fraction of UV bursts are associated with intense Ellerman bombs \citep{Tian2016}.

We used the O {\footnotesize IV} 1399.77/1401.16 \AA~line pair to estimate the electron density of the reconnection region. The intensity ratio of these two spectral lines is sensitive to electron density in the range of $10^{10}$ cm$^{-3}$ to $10^{12}$ cm$^{-3}$. The theoretical relationship between line ratio and electron density was obtained using version 9.0 of the CHIANTI database (Dere et al. 2019) (Fig. 4). From the IRIS spectrum shown in Fig. 3d, we obtained an intensity ratio of $0.23\pm0.05$ for the two O {\footnotesize IV} lines. From the observed ratio we obtained a density of the order of $10^{10.4}$ cm$^{-3}$ (Fig. 4), placing the reconnection site in the chromosphere. 

Analysis of the magnetic field topology in active regions may help us understand the production of WLFs (Song \& Tian 2018; Song et al. 2018b). From the photospheric magnetogram obtained with HMI at 15:12 UT, we reconstructed the three-dimensional (3D) magnetic field structure about 10 minutes before the flare. The vector magnetic field data from the series of hmi.sharp\_cea\_720s were employed for the extrapolation. The lower solar atmosphere is likely not to be force-free, contrary to the force-free assumption widely used  for most magnetic field extrapolations. Here we reconstructed the 3D magnetic field structure using a magnetohydrostatic (MHS) model (Zhu et al. 2013, 2016). This model extrapolates the magnetic field by a magneto-hydrodynamic relaxation approach. It is particularly appropriate for layers where the plasma $\beta$ is relatively high and the force-free assumption fails such as in our case. We performed the extrapolation in a box of 352$\times$320$\times$193 grid points with a resolution element of $0.5^{\prime\prime}$ ($\sim$360 km). From the reconstructed 3D magnetic field, we then calculate the electron current density (\textit{\textbf{J}}) from the following equation:
\begin{equation}
\emph{$\textbf{J}=\displaystyle\frac{1}{\mu_0}(\triangledown \times \textbf{B})$}\label{equation1} ~~~,
\end{equation}
where \textit{\textbf{B}} and $\mu_0$  are the magnetic field strength and magnetic permeability in the vacuum, respectively.

A U-shaped magnetic field structure, the bottom of which touches the photosphere, was seen in the flare region (Fig.5a, b). Such magnetic field structures often result from interaction between emerging magnetic fields and convection flows (Pariat et al. 2004; Xu et al. 2010; Danilovic 2017), and magnetic reconnection between the oppositely directed magnetic field lines at the two sides of a U-shaped structure could lead to the generation of UV bursts and other similar events (Georgoulis et al. 2004; Peter et al. 2014; Chen et al. 2019a, Chen et al. 2019b). In our case, the two sides of the U-shaped structure are the opposite-polarity magnetic fields of the two small sunspots. The center of white-light enhancement (indicated by the cyan dashed line in Fig. 5b, c) is cospatial with the U-shaped structure. From the reconstructed magnetic field configuration we have calculated the spatial distribution of the electric current density (normalized to the magnetic field strength, $\left| J \right| / \left| B \right|$, see Inoue et al. 2018), which appears to concentrate in the lower part of the U-shaped structure (Fig. 5c). These results suggest that the WLF is directly powered by magnetic reconnection in the lower part of the U-shaped structure (less than $\sim$2 Mm above the photospheric height of $\tau$=1).

It is worth mentioning that one side of a U-shaped structure may reconnect with the surrounding magnetic field during flux emergence (e.g., Syntelis et al. 2019; Chen et al. 2019a). In principle this process could produce a flare. Disappearance of pores and weakening of blue shifts may also be expected if the emergence of the U-shaped structure is completed afterwards. However, this scenario appears to be difficult to explain the observed coincidence between the PIL and the WLF, and thus is not favored by our observations.


\section{Summary}

We reported a small WLF, driven by cancellation of opposite-polarity photospheric magnetic fluxes. This compact flare revealed spectral features characteristic of UV bursts but the associated increase of hard X-ray flux is quite marginal. An examination of the magnetic field topology suggests that this flare lies right at the site where a U-shaped magnetic field structure presents.

Our observations appear to be in contradiction with the commonly accepted scenario of WLFs, where electrons are accelerated by magnetic reconnection in the corona and subsequently propagate downward to heat the lower solar atmosphere. Instead, our results provide strong evidence to the hypothesis that some WLFs are produced through energy release by magnetic reconnection in the lower atmosphere. The lower chromosphere or even the photosphere could be locally heated to generate enhanced white-light emission, either through direct heating of the ambient plasma by the reconnection or through non-thermal electrons produced by the reconnection. In the latter case, energies of non-thermal electrons do not need to be as high as needed in the commonly accepted scenario of WLFs, because these electrons are generated in the lower atmosphere and they do not need to pay for the energy loss of traveling a long distance from the corona to the chromospehre/photosphere. Our observations suggest that low-height magnetic reconnection may explain the generation of some type II WLFs and the increasing number of observed small WLFs. 


\acknowledgements

This work is supported by NSFC grants 11803002, 11825301, 11973056, 11790304(11790300) and U1531247, CAS Key Laboratory of Solar Activity (No: KLSA201810, National Astronomical Observatories of CAS), Strategic Priority Research Program of CAS (grant XDA17040507), and the Max Planck Partner Group program. The authors thank the SDO, IRIS, Fermi and GOES teams for providing the data, and Prof. Mingde Ding and Dr. Xiaoli Yan for helpful discussion. IRIS is a NASA Small Explorer mission developed and operated by LMSAL with mission operations executed at NASA Ames Research center and major contributions to downlink communications funded by ESA and the Norwegian Space Center. SDO is a space mission in the Living With a Star Program of NASA.





\newpage

\begin{figure}[!ht]
\centerline{\includegraphics[width=0.9\textwidth]{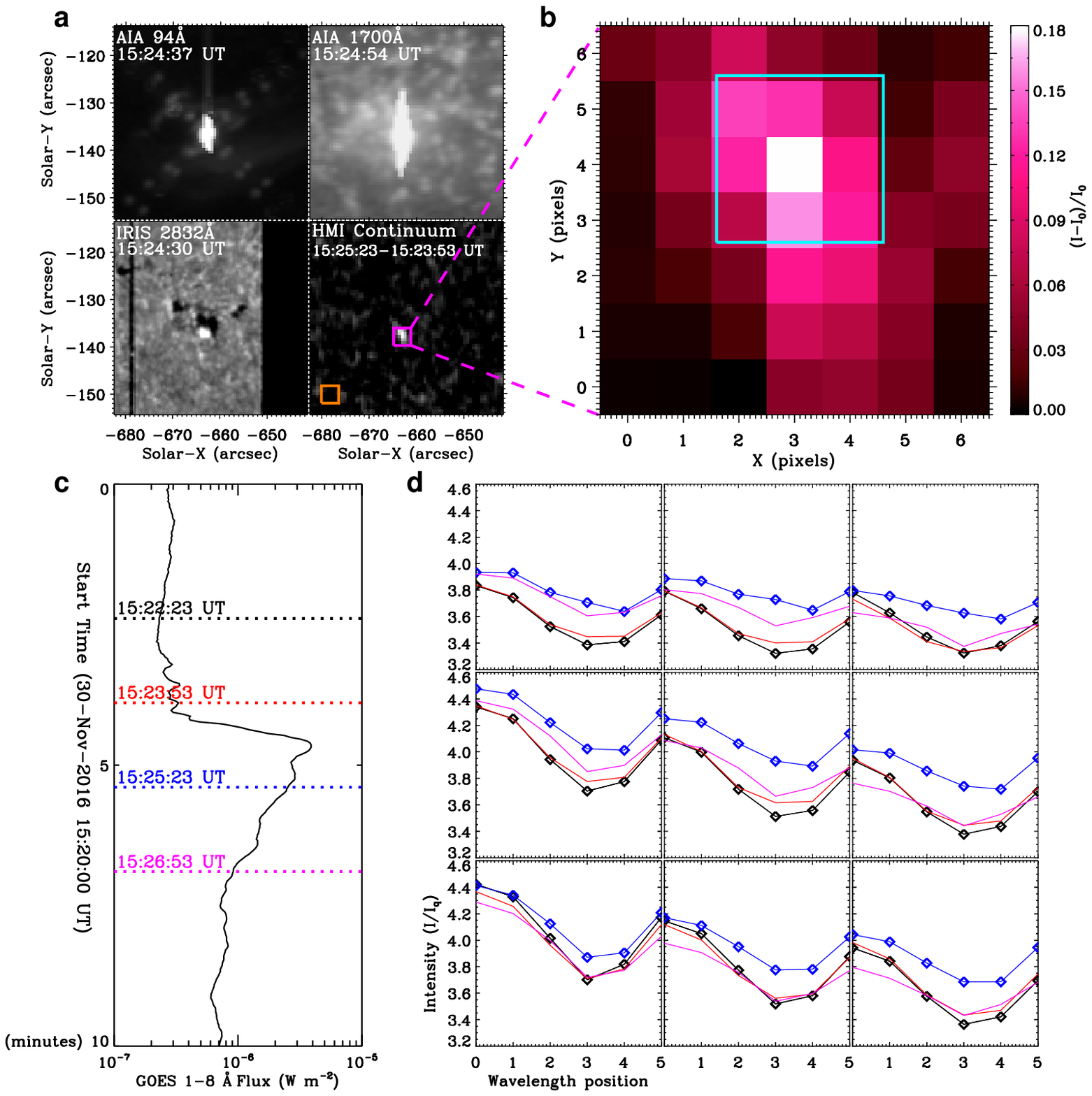}}
\caption{ Identification of the white-light flare. (a) Observations of the flare in the passbands of AIA 94 \AA, 1700 \AA, IRIS 2832 \AA~and HMI visible continuum around 15:24 UT.  For the HMI continuum a difference of two images taken at different times is shown. The purple box shows the region of white-light enhancement (field of view in panel (b)). The brown box marks a quiet region, where the average continuum intensity ($I_q$) is used for normalization in panel (d). (b) White-light enhancement calculated from $(I-I_0)/I_0$, where $I$ and $I_0$ are the HMI continuum intensities around (15:25:23 UT) and before (15:23:53 UT) the flare peak, respectively. (c) Temporal evolution of the GOES 1-8 \AA~soft X-ray flux. The dashed lines with different colors indicate different times. (d) Spectral profiles of Fe I 6173.3 \AA~ measured by HMI at the four different times indicated in panel (c) and at the nine spatial pixels within the cyan box in panel (b). An online animation is available.}
\end{figure}

\begin{figure}[!ht]
\centerline{\includegraphics[width=0.9\textwidth]{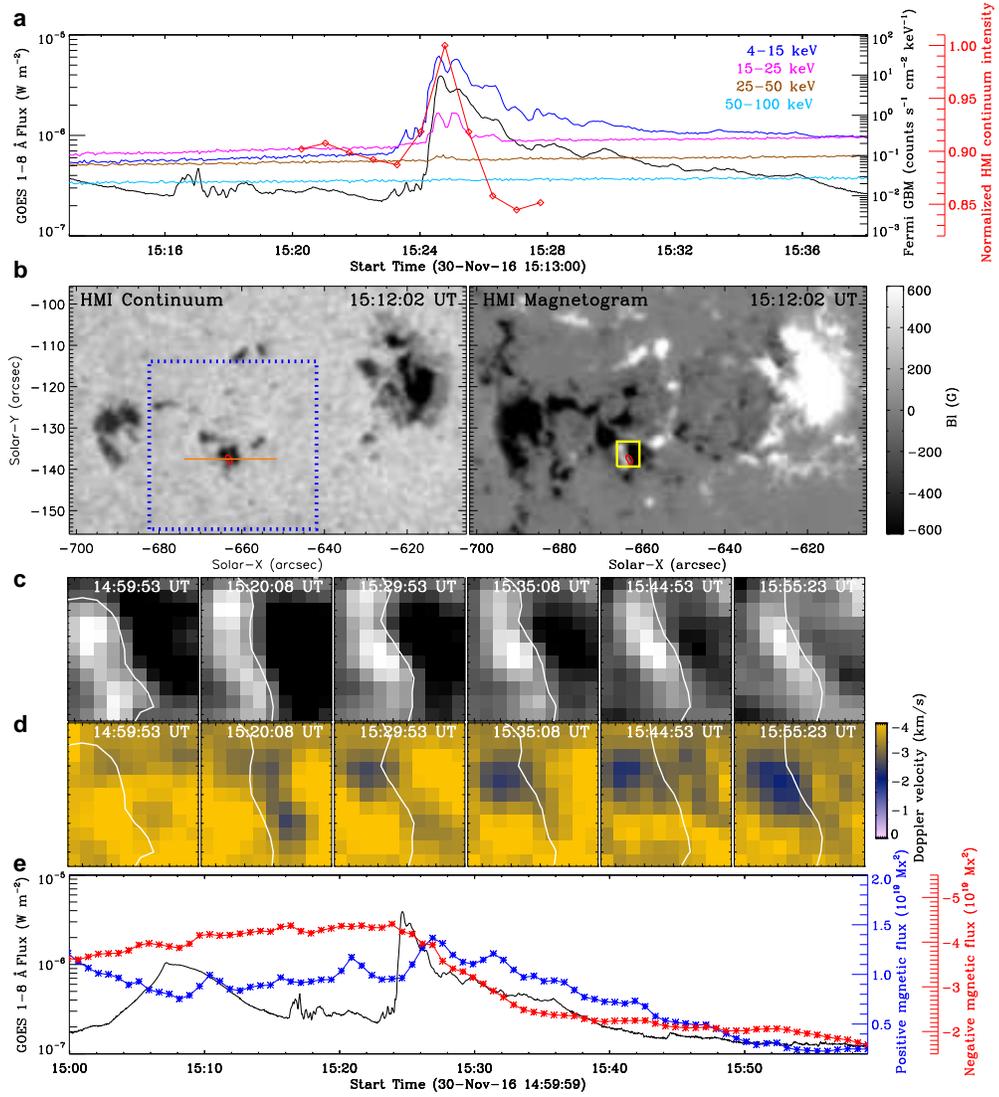}}
\caption{Photospheric magnetic field and X-ray emission associated with the WLF. (a) Temporal evolution of the normalized HMI continuum intensity within the region marked by the purple box in Fig.1a, the X-ray fluxes measured by GOES and Fermi. (b) HMI continuum image and LOS magnetogram taken at 15:12:02 UT (before the flare). The dotted blue box corresponds to the field of view in Fig. 1(a). The red contour indicates the region of significant white-light enhancement ($(I-I_0)/I_0>10\%$). (c)-(d) Sequences of the HMI LOS magnetogram and Doppler velocity in the flaring region marked by the yellow box in panel (b). The white line indicates the polarity inversion line between the two small sunspots. (e) Temporal evolution of the positive (blue) and negative (red) magnetic fluxes, together with the X-ray flux measured by GOES (black). An online animation is available.}
\end{figure}

\begin{figure}[!ht]
\centerline{\includegraphics[width=\textwidth]{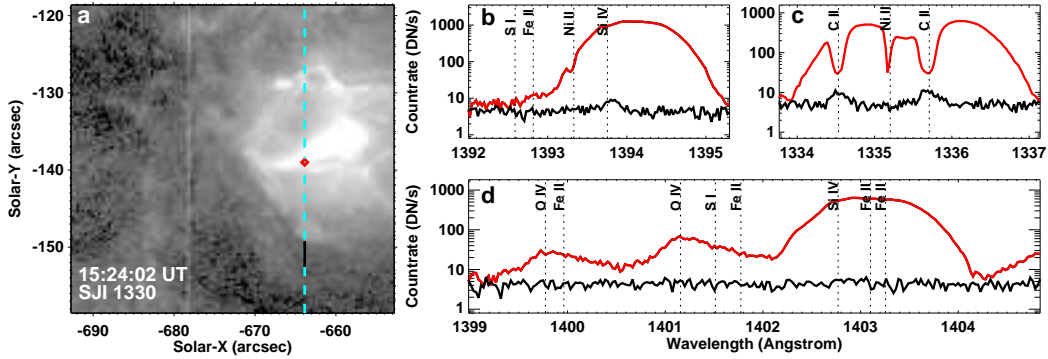}}
\caption{ IRIS observation of the WLF. (a) 1330 \AA~image taken at 15:24:02 UT. The slit location is indicated by the dashed line. (b-d) The red lines represent the spectra taken at a location indicated by the red diamond in panel (a). The black lines show the reference spectra averaged over the section indicated by the black line in panel (a).} 
\end{figure}

\begin{figure}[!ht]
\centerline{\includegraphics[width=0.9\textwidth]{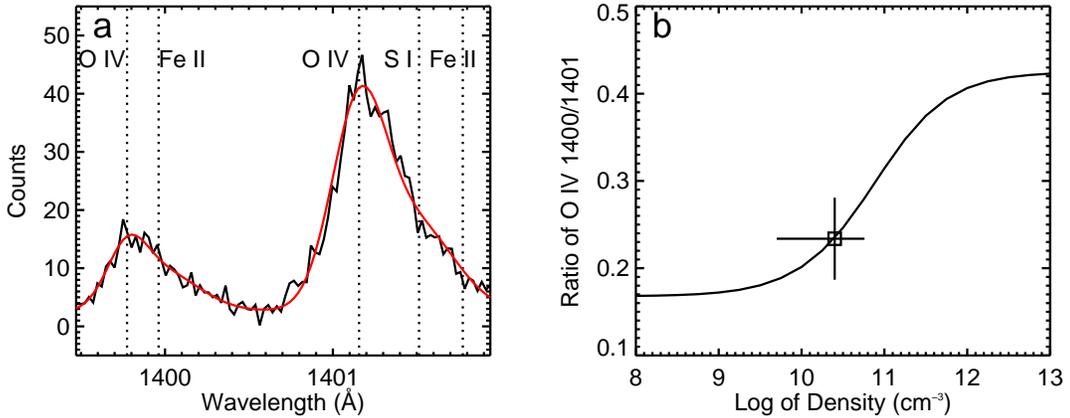}}
\caption{Density diagnostics. (a) Gaussian fitting (red lines) of the O IV 1399.77/1401.16 \AA~line profiles (black lines) observed at the location indicated by the red diamond in Fig. 3a at 15:24:02 UT. Because the two O IV lines are blended with three lines (Fig.3), a five-component Gaussian fitting was applied. (b) Theoretical relationship between line ratio and electron density. The square and bars indicate the measured line ratio/electron density and the associated errors. }
\end{figure}

\begin{figure}[!ht]
\centerline{\includegraphics[width=0.9\textwidth]{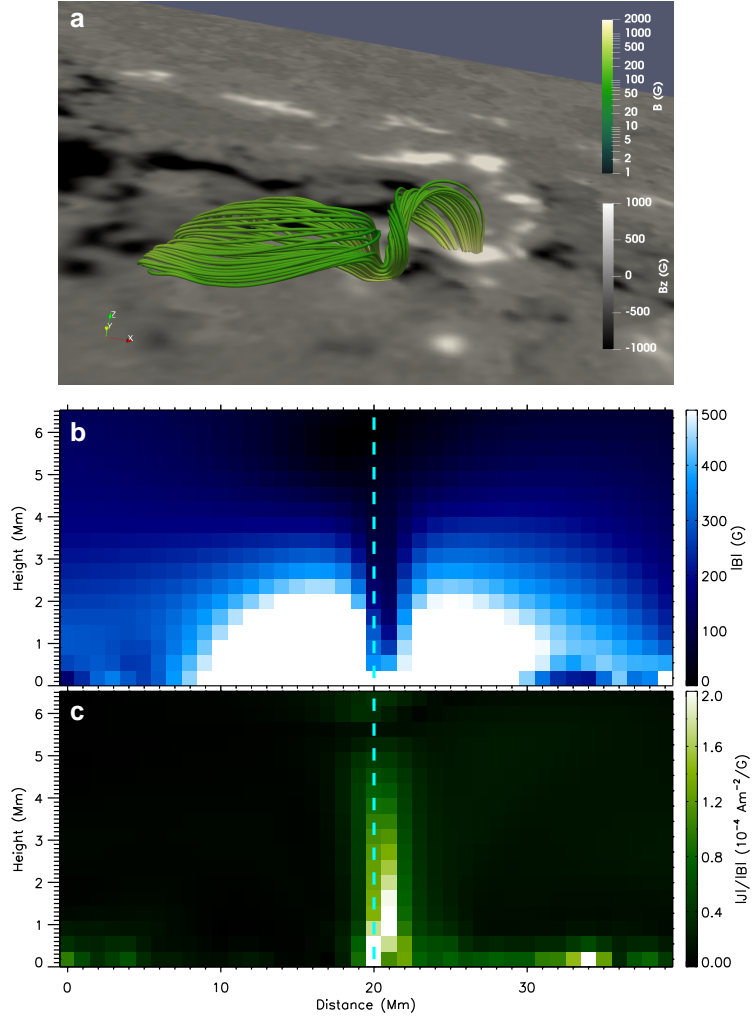}}
\caption{Three-dimensional (3D) magnetic field structure associated with the WLF. (a) 3D magnetic field lines in the flare region obtained with the MHS extrapolation method. A LOS magnetogram taken at 15:12 UT is shown at the bottom. (b-c) Distribution of the magnetic field strength and current density (normalized to the magnetic field strength) on the vertical plane passing through the brown line in figure 2(b). The dashed vertical lines indicate the center of the white-light enhancement region. An online animation is available.}
\end{figure}


\begin{thebibliography}{}

\bibitem[Aboudarham 1986]{Aboudarham1986}
Aboudarham, J. \& Henoux, J. C. 1986, A\&A, 156, 733

\bibitem[Aboudarham 1989]{Aboudarham1989}
Aboudarham, J. \& Henoux, J. C. 1989, Sol. Phys., 121, 19

\bibitem[Atwood 2009]{Atwood2009}
Atwood, W. B. et al. 2009, ApJ, 697, 1071

\bibitem[Cao 2010]{Cao2010}
Cao, W., Gorceix, N., Coulter, R., et al. 2010, Astronomische Nachrichten, 331, 636

\bibitem[Carrington 1859]{Carrington1859}
Carrington, R. C. 1859, MNRAS, 20, 13

\bibitem[Chen 2001]{Chen2001}
Chen, P.-F., Fang, C.,\& Ding, M. D. 2001, ChJA\&A, 1, 176

\bibitem[Chen 2019]{Chen2019a}
Chen, Y.-J., et al. 2019a, ApJL, 875, L30

\bibitem[Chen 2019]{Chen2019b}
Chen, Y.-J., et al. 2019b, Science China Technological Sciences, 62, 1555

\bibitem[Couvidat 2012]{Couvidat2012}
Couvidat, S., et al. 2012, SoPh, 278, 217

\bibitem[Danilovic 2017]{Danilovict2017}
Danilovic, S. 2017, A\&A, 601, A122 

\bibitem[De Pontieu 2014]{De Pontieu2014}
De Pontieu, B. et al. 2014, SoPh, 289, 2733

\bibitem[Dere 2019]{Dere2019}
Dere, K. P., Del Zanna, G., Young, P. R., et al. 2019, ApJS, 241, 22

\bibitem[Ding 1994]{Ding1994}
Ding, M. D., Fang, C., Gan, W. Q., et al. 1994, ApJ, 429, 890 

\bibitem[Fang 1995]{ Fang1995}
Fang, C. \& Ding, M. D. 1995, A\&AS, 110, 99

\bibitem[Fletcher 2007]{ Fletcher2007}
Fletcher, L., Hannah, I. G., Hudson, H. S., et al. 2007, ApJ. 656, 1187

\bibitem[Georgoulis 2002]{Georgoulis2002 }
Georgoulis, M. K., Rust, D. M., Bernasconi, P. N., et al. 2002, ApJ, 575, 506

\bibitem[Hao 2012]{Hao2012}
Hao, Q., Guo, Y., Dai, Y., et al. 2012, A\&A, 544, L17

\bibitem[Hodgson 1859]{Hodgson1859}
Hodgson, R. 1859, MNRAS, 20, 15

\bibitem[Hudson 2006]{Hudson2006}
Hudson, H. S., Wolfson, C.J., Metcalf, T.R. 2006, SoPh, 234, 79

\bibitem[Hudson 2016]{Hudson2016}
Hudson, H. S. 2016, SoPh, 291, 1273

\bibitem[Inoue 2018]{Inoue2018}
Inoue, S., Shiota, D., Bamba, Y. \& Park, S.-H. 2018, ApJ, 867, 83

\bibitem[Jess 2008]{Jess2008}
Jess, D. B., Mathioudakis, M., Crockett, P. J. \& Keenan, F. P. 2008, ApJL, 688, L119

\bibitem[Kahler 1982]{Kahler1982}
Kahler, S. W. 1982, JGR, 87, 3439

\bibitem[Kuhar 2016]{Kuhar2016}
Kuhar, M., Krucker, S., Mart\'inez Oliveros, J.C. et al. 2016, ApJ, 816, 6

\bibitem[Lemen 2012]{Lemen2012}
Lemen, J. R., Title, A. M., Akin, D. J. et al. 2012, SoPh, 275, 17

\bibitem[Machado 1989]{Machado1989}
Machado, M. E., Emslie, A. G., \& Avrett, E. H. 1989, SoPh, 124, 303

\bibitem[Matthews 2003]{ Matthews2003}
Matthews, S. A., van Driel-Gesztelyi, L., Hudson, H. S., et al. 2003, A\&A, 409, 1107

\bibitem[Meegan 2009]{ Meegan2009}
Meegan, C., Lichti, G., Bhat, P. N., et al. 2009, ApJ, 702, 791

\bibitem[Metcalf 2003]{Metcalf2003}
Metcalf, T. R., Alexander, D., Hudson, H. S., \& Longcope, D. W. 2003, ApJ, 595, 483

\bibitem[Neidig 1989]{Neidig1989a}
Neidig, D. F. 1989, SoPh, 121, 261

\bibitem[Pariat 2004]{ Pariat2004}
Pariat, E., Aulanier, G., Schmieder, B. et al. 2004, ApJ, 614, 1099

\bibitem[Pesnell 2012]{Pesnell2012}
Pesnell, W. D.,  Thompson, B. J. \& Chamberlin, P. C. 2012, SoPh, 275, 3

\bibitem[Peter 2014]{ Peter2014}
Peter, H., Tian, H., Curdt, W. et al. 2014, Sci, 346, 1255726

\bibitem[Ryan 1983]{Ryan1983}
Ryan, J. M., Chupp, E. L., Forrest, D. J., et al. 1983, ApJL, 272, L61.

\bibitem[Scherrer 2012]{Scherrer2012}
Scherrer, P. H., Schou, J., Bush, R. I., et al. 2012, SoPh, 275, 207

\bibitem[Schmieder 2004]{Schmieder2004}
Schmieder, B., Rust, D. M., Georgoulis, M. K., et al. 2004, ApJ, 601, 530


\bibitem[{\v{S}}vanda et al.(2018)]{2018ApJ...860..144S} 
{\v{S}}vanda, M., Jur{\v{c}}{\'a}k, J., Ka{\v{s}}parov{\'a}, J., et al.\ 2018, \apj, 860, 144


\bibitem[Song 2018]{Song 2018a}
Song, Y. L., Guo, Y., Tian, H., et al. 2018a, ApJ, 854, 64

\bibitem[Song 2018b]{Song 2018b}
Song, Y. L., Tian, H., Zhang, M., et al. 2018b, A\&A, 613, A69

\bibitem[Song 2018]{Song 2018}
Song, Y. L. \& Tian, H. 2018, ApJ, 867, 159

\bibitem[Sylwester 2000]{Sylwester2000}
Sylwester, B., \& Sylwester, J. 2000, SoPh, 194, 305

\bibitem[Syntelis et al.(2019)]{Syntelis2019} 
Syntelis, P., Priest, E.~R., \& Chitta, L.~P.\ 2019, \apj, 872, 32

\bibitem[Tian et al. 2016]{Tian2016}
Tian, H., Xu, Z., He, J., Madsen, C. 2016, ApJ, 824, 96

\bibitem[Tian et al. 2018]{Tian2018}
Tian, H., Zhu, X. S., Peter, H., et al. 2018, ApJ, 854, 174


\bibitem[Watanabe 2010]{Watanabe2010}
Watanabe, K., Krucker, S., Hudson, H., et al. 2010, ApJ, 715, 651

\bibitem[Wang 1993]{Wang1993}
Wang, J. X. \& Shi, Z. X. 1993, SoPh, 143, 119

\bibitem[Xu 2010]{Xu2010 }
Xu, Z., Lagg, A., \& Solanki, S. K. 2010, A\&A, 520, A77


\bibitem[Yan 2016]{ Yan 2016}
Yan, X. L., Priest, E. R, Guo, Q. L, et al. 2016, ApJ, 832, 23

\bibitem[Young 2018]{ Young2018}
Young, P. R., Tian, H., Peter, H., et al. 2018, Space Sci. Rev., 214, 120 

\bibitem[Yurchyshyn 2017]{ Yurchyshyn2017}
Yurchyshyn, V., Kumar, P., Abramenko, V., et al. 2017, ApJ, 838, 32

\bibitem[Zhu 2013]{Zhu2013}
Zhu, X. S., Wang, H. N., Du, Z. L., \& Fan, Y. L. 2013, ApJ, 768, 119

\bibitem[Zhu 2016]{Zhu2016}
Zhu, X. S., Wang, H. N., Du, Z. L., \& He, H. 2016, ApJ, 826, 51


\end{thebibliography}
\end{document}